\documentstyle [emulateapj]{article}

\begin{document}

\title{Observations of OJ 287 from the Geodetic VLBI Archive of the Washington
Correlator}

\author{C. E. Tateyama}
\affil{Centro de R\'adio-Astronomia e Aplica\c c\~oes Espaciais/CRAAE-INPE,
Instituto Presbiteriano Mackenzie, Rua da Consola\c c\~ao 896,
01302-000, S\~ao Paulo, SP, Brazil}

\author{K. A. Kingham}
\affil{U.S. Naval Observatory, Earth Orientation Dept,
3450 Massachusetts Ave, Washington D.C. 20392, USA}

\author{P. Kaufmann}
\affil{Centro de R\'adio-Astronomia e Aplica\c c\~oes 
Espaciais/CRAAE-Mackenzie,
Instituto Presbiteriano Mackenzie, Rua da Consola\c c\~ao 896,
01302-000, S\~ao Paulo, SP, Brazil}

\author{B. G. Piner}
\affil{ Jet Propulsion Laboratory, California Institute of Technology, 
MS 238-332, 4800 Oak Grove Dr., Pasadena, CA 91109}

\author{L. C. L. Botti}
\affil{Centro de R\'adio-Astronomia e Aplica\c c\~oes Espaciais/CRAAE-INPE,
Instituto Presbiteriano Mackenzie, Rua da Consola\c c\~ao 896,
01302-000, S\~ao Paulo, SP, Brazil}

\and

\author{A. M. P. de Lucena}
\affil{ROEN, R\'adio Observat\'orio Espacial do Nordeste, CRAAE-INPE,
Estrada do Fio, 6000, Eus\'ebio (Fortaleza), CE, Brazil} 

\begin{abstract}

We present 27 geodetic VLBI maps of OJ 287 obtained from the archive of 
the Washington correlator.  The observations presented here were made 
between 1990 October and 1996 December. During this period  
a sequence of six superluminal components has been identified.
We measured the proper motion of these components to be approximately 0.5 mas 
yr$^{-1}$, which is
about twice as high as that seen in previous VLBI observations.
These results imply a higher component ejection rate than previously observed,
in good agreement with the observed occurrences of radio outbursts.
We have examined a possible connection between VLBI components
and optical flares in the framework of a binary black hole system.

\end{abstract} 


\section{Introduction} 

OJ 287 is a highly variable radio source at
z = 0.306 (Miller, French, \& Hawley 1978;  Sitko \& 
Junkkarinen 1985). The nature of the underlying galaxy 
is not well known; Kinman (1975) and Hutchings et al. (1984) 
detected a nebulosity that might be a galaxy. 
The light curve of OJ 287 presents a complex structure, but 
its most interesting aspect is the series of prominent flares 
which recur with a period of 12 years (Sillanp\"a\"a et al. 1988). 
Recent V-band observations  have revealed double-peaked outbursts
(Sillanp\"a\"a et al. 1996;  Lehto \& Valtonen 1996).
To reproduce the cyclic 12 year optical flares,
OJ 287 has become the best candidate to harbor a supermassive binary
black hole system.  Sillanp\"a\"a et al. (1988) used
such a system in which the light variations were related to tidally
induced mass flows from the accretion disk into the black hole to explain
the periodicity in OJ 287's light curve.
The orbit of the secondary black hole was assumed to be coplanar
with the accretion disk. Lehto \& Valtonen (1996) suggested
a binary system model where the primary is surrounded 
by an accretion disk with a
high inclination angle relative to the orbit of the secondary. 
A major flare would be observed whenever the secondary crossed the disk
of the primary.  
The binary system model of Villata et al. (1998) with two independent 
relativistic jets  and 
the nodding disk model of Katz (1997) use a sweeping beam
approach to explain the same optical flares.

The radio light curves are dominated by outbursts at irregular intervals of 
about
a year,  and modulated by a much longer non-periodic timescale (Usher 1979).
Analysis of radio-optical events in OJ 287 has shown correlated activity.
Usher (1979) compared
events from 1966 to 1978 and proposed that most optical
outbursts are synchronous with radio counterparts. 
Valtaoja, Sillanp\"a\"a, \& Valtaoja (1987) studied isolated flaring events
and found that optical variations preceded the radio ones
by a few months. 

VLBI observations of OJ 287 have been made at 5 GHz 
(Roberts, Gabuzda, \& Wardle 1987; 
Gabuzda, Wardle, \& Roberts 1989), 8 GHz (Vicente, Charlot, \& Sol 1996), 
and 43 and 100 GHz (Tateyama et al. 1996).
Superluminal motion was first detected from 
VLBI polarization data at 5 GHz (Roberts et al. 1987). 
Previously to this paper, three superluminal knots --- named K1, K2 and K3 ---
have been followed from 1981 to 1988.
The observed proper motion of these components 
corresponds to an apparent superluminal speed between 
3.7 and 5.1$c$ (H$_0$ = 65 km s$^{-1}$ Mpc$^{-1}$, q$_0$ =  0.5). 
In a single epoch VLBI observation at 8.3 GHz, a new jet component K4 and 
another   
polarization component located between the core and K4 named K5 have been added 
to the list
of components of  OJ 287 by Gabuzda \& Cawthorne (1996).
More recently, Vicente et al. (1996) fitted a 
helical path with two consecutive loops to the trajectory of component K3.

We present 27 geodetic VLBI maps at 8.3 GHz obtained from the
Washington correlator's database.
Geodetic VLBI maps have been successfully used for astrophysical
studies of compact sources (e.g., Guoquiang, R\"onn\"ang \& B\aa \aa th 1987;  
Charlot 1990;  Britzen et al. 1994;  Vicente et al. 1996). More recently, 
Piner \& Kingham (1998), and Tateyama et al. (1998)
have obtained interesting results for a group of EGRET blazars 
and BL Lac respectively using the Washington VLBI correlator's database.

\section{Observations}

The VLBI observations used in this paper are from geodetic VLBI 
data  archived at the Washington correlator. 
These data were obtained from geodetic     
dual-frequency VLBI experiments (Rogers et al. 1983) carried out by the  
Naval Observatory (Eubanks et al. 1991), the National Oceanographic and 
Atmospheric Administration (Carter, Robertson, \& MacKay 1985), the  Crustal 
Dynamics 
Project, and the Space Geodesy Project, (Coates et al. 1985; Smith \&     
Turcotte 1993). The VLBI observations were processed at the Washington VLBI 
Correlator at the U.S. Naval Observatory (USNO). 

All observations were obtained with Mark-III dual-frequency VLBI
receivers at both X- and S-band (centered at 8.5 and 2.3 GHz respectively), 
providing noise temperatures of 70 K - 200 K.  X-band consists of  8 
individual
channels of 2 MHz bandwidth spanning the range 8.2 to 8.9 GHz. All stations 
were equipped with H-masers as the local frequency standards.
Geodetic VLBI data useful for imaging OJ 287 begins to appear in the
Washington Correlator archive around 1990.
Table 1 lists all
observations used in this work: column 1 gives the
epochs of the observations, column 2 the names of the experiments, 
column 3 the
antennas participating in the experiments, column 4 the peak brightness of the 
images,
column 5 the interferometric dirty beams, 
column 6 the position angles of the beams measured from north through east,
and the last column the number of scan-baselines in the observations, 
where the number of scan-baselines
is the sum over all the scans of the number of baselines per scan. 
The dirty beams were about 0.5 - 0.6 mas in size for all maps.
We have used a restored circular beam of 0.6 mas on the maps.
The dynamic range defined as the ratio of the peak flux
per beam to the lowest positive contour in the maps is about 300:1. 
The total length in which the moving components in OJ 287 can be followed with
the geodetic network appears to be limited to radii
not much greater than 1.5 mas.  This is about half the distance
reached by the knots in the 5 GHz VLBI observations made by Roberts et al. 
(1987) and Gabuzda et al. (1989).
Except for CRD96GH (Table 1) where the resulting map is the sum of the
experiments CRD96G and CRD96H observed on Sep 19 and Sep 20 of 1996
respectively, all maps were obtained from a single geodetic observation.

\begin{deluxetable} {cllccrr}
\tablecaption{8.4 GHz VLBI observations of OJ 287. }
\tablehead{
\colhead{Epoch} & \colhead{Name} & \colhead{Antennas} & \colhead{Peak} &
\colhead{Synthesized } & \colhead{Beam angle} & \colhead{Number of} \\
& & & \colhead{brightness} & \colhead{beam} &  & \colhead{scan-baselines} \\
& & & \colhead{[Jy beam$^{-1}$]} & \colhead{[mas x mas]}  & \colhead{[deg]} 
}
\startdata 
90 Oct  27 & NAPSA    & JYEAC &  4.9  & 1.6  x  1.3  &  39  & 139 \\
92 Aug  03 & IR752      & GJREV   &  2.3   &  1.8  x 0.7   & -19    &   48 \\ 
92 Aug  10 & IR753      & GJREV    &  2.4     &  1.0  x 0.7   &  -15   &   53  
\\ 
93 Jun 24 & NB02        & AIWVGT  &1.3    &  1.6  x 0.6   & -2     &  66  \\ 
93 Dec 10 & NB08       & GIWTV     & 1.5    &  2.5   x 0.5  & -10    &  63  \\
94 Sep 02 & NB16       & GXAIWTV  & 1.0   &  1.5   x 0.6  &  2       & 59  \\
94 Dec 13 & NE85       & FAKWVN    & 1.0  &  0.5   x 0.5  & -35      & 63   \\ 
95 Jan 25 & NA91       & FAKWV     & 0.9   &  0.7   x 0.4  & 20        & 22    
\\ 
95 Apr 18 & NA103     & FAKWVO   & 0.8   &  0.9   x 0.6  & -21     & 51    \\ 
95 Jun 14 & NXS6       & GAMWOV    & 0.8  &  1.1   x 0.7  & -29      & 51   \\
95 Sep 19 & NA125    & GFAKWNV   & 1.2  &  1.3   x 0.6  & -4       & 29   \\ 
95 Oct 17 & NA129    & GFAKHLNV   & 1.4 &   0.7  x 0.5   & 17        & 80   \\
95 Dec 12 & NA137    & FAKNOV      & 2.0  &   0.7  x 0.6   & -75    & 63  \\ 
96 Mar 05 & NA149    & FAKNOV   & 1.2    &  0.8    x 0.6   & -39     & 62    \\ 
96 Mar 26 & NA152    & FAKNOV   & 1.1    &  0.8    x 0.6   & -34   & 61    \\ 
96 Apr 02 & NA153    & FAKMNV    & 1.2   &  0.9    x 0.6   & 2     & 67    \\ 
96 Apr 24 & NXS10     & GAKNOV    & 0.9   &  0.9   x 0.7   & -14   & 22    \\ 
96 Aug 20 & NA173     & FAKWNOV  & 1.1  &  0.6   x  0.5 &   15   & 63   \\
96 Sep 03 & NA175     & FAKNOV    & 1.1   &  0.8   x 0.5   & -58   & 51    \\ 
96 Sep 17 & NA177     & FAKNOV    & 1.1   &  0.7   x 0.6   & -27   & 69    \\
96 Sep 19 & CRD96GH  & AKNOV     & 0.9   &  0.8   x 0.6    & -22  & 204   \\ 
96 Oct 01 & NA179     & FAKNOV    & 1.2    &  1.0  x 0.5    & -16   & 68    \\
96 Oct 02 & CRD96J   & AKNOV       & 1.0   &   0.9  x 0.5   &  -18   & 98   \\
96 Oct 03 & CRD96K   & AKNOV       & 0.9   &  1.0   x 0.5   & -20  & 79    \\
96 Oct 04 & CRD96L   & AKNOV       & 1.0   &   1.0  x 0.6   & -23    & 91   \\
96 Nov 12 & NA185    & FAKNOV      & 1.3   &   0.6  x 0.4    &  8       & 61   
\\
96 Dec 10  & NA189   & FAKNOV      & 1.1  &   0.9  x 0.5    & -46    & 62   \\ 
\hline
\enddata

\tablecomments{
A = Gilcreek (Alaska, 26 m),  
C = Hat Creek (California, 26 m),
E = Westford (Massachusetts, 18 m),
F = Fortaleza (Brazil, 14 m),  
G = Algopark (Ontario, 46 m), 
H = MK-VLBA (Hawaii, 25 m)
I = Matera (Italy, 20 m),
J = Mojave (California, 12 m),
K = Kokee (Hawaii, 20 m),  
L = NL-VLBA (Iowa, 25 m)
M = Miami20 (Florida, 20 m),
N = NRAO20 (West Virginia, 20 m), 
O = Ny Alesund (Norway, 20 m), 
R = Richmond (Florida, 18 m),
T = Onsala (Sweden, 20 m),
V = Wettzell (Germany, 20 m), 
X = Ylow7296 (Northwest Territories, 10 m), 
Y = Plattville (North Carolina, 5 m),
W= NRAO85 (West Virginia, 26 m)}

\end{deluxetable}

The data were coherently averaged to four seconds to determine
the visibilities. The data were calibrated and fringed using standard
routines from AIPS software package, and the images were produced using the
self-calibration procedures (e.g. Pearson \& Readhead 1984) of the Caltech
Difmap software package.  

\section{Results}

Throughout this paper we assume a standard Friedmann cosmology
with H$_0$ = 65 km s$^{-1}$ Mpc$^{-1}$ and q$_0$ =  0.5. At the distance of 
OJ 287 (z = 0.306), an angular size of 1 mas corresponds to a linear distance
of  4.3 pc.
The 8 GHz radio maps are presented in Figure 1. The source shows a
core-jet structure, with jet components moving in a westerly direction.
Table 2 lists the flux,
position, and size of each  component obtained by model fitting 
the observed visibilities with elliptical Gaussian components. 
Position angle is measured from north through east. 

\begin{figure*}
\plotone{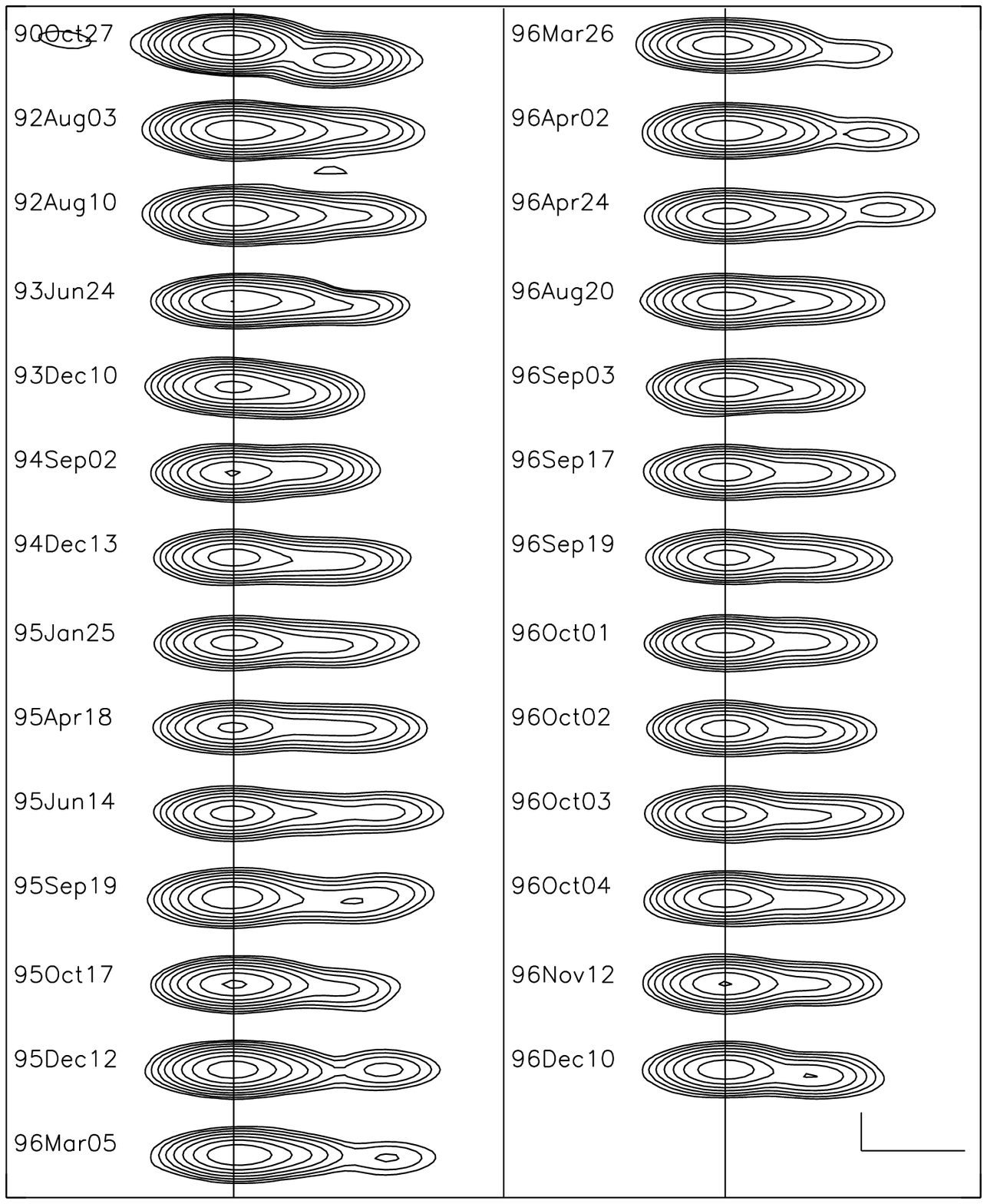}
\caption{8 GHz VLBI maps of OJ 287. Contour levels increase 
as 2$^{n}$.  The lowest contour level is 0.01 Jy beam$^{-1}$,
corresponding to $n=0$.
The restored circular beam is 0.6 mas. 
The stationary component (the VLBI core) is on the east
side (left), while the moving components are to the west (right).
The tick in the lower part of the map indicates the size in mas.
Note that the maps are stretched in right ascension.}
\end{figure*}

\begin{deluxetable} {cccrrccr}
\tablecaption{Modelfits to OJ 287 at 8 GHz.}
\tablehead{
\colhead{Epoch} & \colhead{Comp.} & \colhead{Flux} & \colhead{r} &
\colhead{PA\tablenotemark{a}} & \colhead{$\theta$\tablenotemark{b}} 
&   \colhead{AR\tablenotemark{c}} & \colhead{PA\tablenotemark{d}}\\
& & \colhead{[Jy]} & \colhead{[mas]} & \colhead{[$^{\circ}$]} & \colhead{[mas]} 
& &   \colhead{[$^{\circ}$]}
}
\startdata
90 Oct  27 & core & 4.40 & 0 & ... & 0.15 & 0.8 & 0 \\
& C1 & 1.21 & 0.98 & -109 & 0.70 & 0.8 & 60 \\
92 Aug  03 & core & 2.20 & 0  & ... & 0.19 & 0.8 & 80 \\ 
& C3 & 0.76  & 0.51  & -98 & 0.30 & 0.8 & -38 \\ 
& C2 & 0.21 & 1.24  & -110  & 1.10 & 0.8 & 5  \\ 
92 Aug  10 & core & 2.30  & 0 & ...  & 0.17 & 0.8 & -60 \\ 
& C3 & 0.55  & 0.49  & -99 & 0.30 & 0.8 & -45 \\ 
& C2 & 0.27 & 1.17   & -110 & 0.90 & 0.7 & 50\\ 
93 Jun 24 & core & 1.05  & 0  & ... & 0.15 & 0.7 & 0\\ 
& C4 & 0.61 & 0.39  & -89 & 0.54 & 0.6 & 105 \\ 
& C3 & 0.23 & 0.98  & -99 & 1.10 & 0.7 & -70\\ 
93 Dec 10 & core & 1.56  & 0 & ... & 0.19 & 1.0 & ... \\ 
& C4 & 0.54 & 0.55 & -102 & 0.30 & 0.7 & -10  \\ 
& C3 & 0.48 & 1.19 & -85  &  0.80 & 0.5 & 0     \\
94 Sep 02 & core & 1.25 & 0 & ... & 0.10 & 1.0 & ... \\ 
& C5 & 0.55 & 0.62 & -88 & 0.40 & 1.0  & ... \\  
94 Dec 13 & core & 0.93 & 0  & ... & 0.13 & 0.8 & -10 \\ 
& C5 & 0.89 & 0.79  & -95 & 0.80  & 0.8 & 90 \\  
95 Jan 25 & core & 0.97  & 0 & ... & 0.17 & 0.9 & -75  \\ 
& C5 & 0.54 & 0.82  & -93 & 0.70 & 0.8 & -90  \\  
95 Apr 18  & core & 0.76  & 0 & ... & 0.17 & 0.8 & -80 \\ 
& C5 & 0.58 & 0.98 & -90 & 0.90 & 0.8 & 80 \\  
95 Jun 14 & core & 0.83 & 0 & ... & 0.17 & 0.7 & -41 \\ 
& C5 & 0.28 & 1.06  & -99 & 1.16 & 0.7 & 10  \\ 
95 Sep 19 & core & 1.26 & 0 & ... & 0.20 & 0.8 & -50  \\ 
& C5 & 0.40 & 1.13 & -96 & 0.93 & 0.8 & -8  \\  
95 Oct 17 & core & 1.24 & 0 & ... & 0.12 & 0.7 & 50 \\ 
& C6 & 0.36 & 0.30 & -86 & 0.35 & 1.0 & ... \\ 
& C5 & 0.18  & 1.13 & -98 & 1.08  & 1.0 & ... \\ 
95 Dec 12 & core & 1.92 & 0 & ... & 0.11 & 0.8 & -50  \\ 
& C6 & 0.24 & 0.36  & -97 & 0.25 & 0.8 & -56  \\ 
& C5 & 0.28  & 1.30 &  -90 & 1.30 & 0.9 & -86  \\ 
96 Mar 05 & core & 1.18  & 0 & ... & 0.17 & 0.8 & -73 \\ 
& C6 & 0.24 & 0.43  & -89 & 0.35  & 0.9 & -10 \\
& C5 & 0.12 & 1.35  & -99 & 1.13  & 0.9 & 60  \\
96 Mar 26  & core & 0.95 & 0 & ... & 0.17 & 0.7 & -70\\
& C6 & 0.35 & 0.38 & -95 & 0.38  & 0.9 & -1  \\ 
& C5 & 0.13 & 1.33 & -108 & 1.20 & 0.8 & 3  \\ 
96 Apr 02 & core & 1.12 & 0 & ... & 0.17 & 0.8 & -75 \\ 
& C6 & 0.30 & 0.42 & -90 & 0.25  & 0.7 & 50 \\ 
& C5 & 0.03 & 1.39 & -100 & 1.12 & 0.6 & 10  \\ 
96 Apr 24 & core & 0.92 & 0 & ... & 0.15 & 0.8 & 0\\ 
& C6 & 0.24 & 0.59 & -90  & 0.43 & 0.8 & -80 \\ 
& C5 & 0.05 & 1.42 & -99 & 1.49 & 0.8 & -63  \\ 
96 Aug 20 & core & 1.06 & 0 & ... & 0.12 & 0.7 & 80  \\ 
& C6 & 0.45  & 0.70 & -89 & 0.50 & 0.8 & 85  \\ 
96 Sep 03  & core & 0.93  & 0 & ... & 0.12 & 0.8 & 81 \\ 
& C6 & 0.42 & 0.73 & -95 & 0.42 & 0.7 & -66  \\ 
96  Sep 17 & core & 1.03 & 0 & ... & 0.15 & 0.7 & 62 \\ 
& C6 & 0.35 & 0.78 & -92 & 0.43 & 0.8 & 84  \\
96 Sep 19 & core & 0.91 & 0 & ... & 0.17 & 0.7 & 3 \\
& C6 & 0.33 & 0.80 & -92 & 0.40 & 0.8 & 80  \\ 
96 Oct 01 & core & 1.24 & 0 & ... & 0.18 & 0.8 & 18\\ 
& C6  & 0.32 & 0.79 & -86 & 0.40 & 0.7 & 56 \\ 
\enddata
\end{deluxetable}

\begin{deluxetable} {cccrrccr}
\tablenum{2--Continued}
\tablehead{
\colhead{Epoch} & \colhead{Comp.} & \colhead{Flux} & \colhead{r} &
\colhead{PA\tablenotemark{a}} & \colhead{$\theta$\tablenotemark{b}} 
& \colhead{AR\tablenotemark{c}} & \colhead{PA\tablenotemark{d}}\\
& & \colhead{[Jy]} & \colhead{[mas]} & \colhead{[$^{\circ}$]} & \colhead{[mas]} 
& &  \colhead{[$^{\circ}$]}
}
\startdata
96 Oct 02 & core  & 0.99 & 0 & ... & 0.17 & 0.8 & -12 \\ 
& C6 & 0.27 & 0.82 & -96 & 0.48 & 0.6 & -59 \\ 
96 Oct 03 & core & 0.89 & 0 & ... & 0.16 & 0.8 & -18 \\ 
& C6 & 0.43 & 0.84 & -97 & 0.85 & 0.7 & -80 \\  
96 Oct 04 & core & 1.03 & 0 & ... & 0.16 & 0.8 & -6 \\ 
& C6 & 0.37 & 0.84 & -92 & 0.50 & 0.8 & -85  \\ 
96 Nov 12 & core & 1.41  & 0  & ... & 0.17 & 0.8 & 20 \\ 
& C6 & 0.43 & 0.87  & -91 & 0.66 & 0.6 & 84 \\ 
96 Dec 10 & core & 1.16 & 0 & ... & 0.18 & 0.8 & -135 \\
& C6 & 0.49 & 0.86 & -98 & 0.50 & 0.6 & 70 \\ 
\enddata
\tablenotetext{a} {Position angle of the component.}
\tablenotetext{b} {Major axis of the elliptical component.}
\tablenotetext{c} {Axial ratio of the elliptical component.}
\tablenotetext{d} {Position angle of the major axis of the elliptical 
component.}
\end{deluxetable}

The motion of the components away from the core
can be better appreciated by inspecting Figure 2,  where the
separation of components from the core is plotted against time.
The straight lines provide estimates of the
presumed zero separation of the components C1 to C6 under the assumption
of rectilinear motion at constant velocity.
The measured proper motions of the components  C1 to C6 are
0.74 $\pm$ 0.40,  0.44 $\pm$ 0.05, 
0.52 $\pm$ 0.09,  0.40 $\pm$ 0.13, 
0.46 $\pm$ 0.05, and 0.58 $\pm$ 0.07 mas yr$^{-1}$ respectively, corresponding 
to
an average superluminal speed of approximately 9$c$. 
These values are almost twice as high as those seen 
in VLBI observations made at 5 GHz by Roberts et al. (1987) 
and Gabuzda et al. (1989). 
A similar proper motion has also been obtained by Vicente et al. (1996) 
for the fastest part of the motion of component K3  (0.4 mas yr$^{-1}$).
The number of components observed over this time range, together with their
measured speeds, suggests more frequent ejection of VLBI components than
has been previously estimated for this source by Vicente et al. (1996), who 
concluded that
VLBI component ejections occurred at intervals of one-half the orbital period of 
the
putative binary black hole system, or once every six years.
The small separation between successive components  obtained by 
Gabuzda \& Cawthorne (1996) at 8 GHz and Tateyama et al. (1996) at 43 GHz
also suggests a higher ejection rate. We can constrain these results using radio 
light curves, 
provided that radio outbursts are connected with VLBI components.

\begin{figure*}[!t]
\plotone{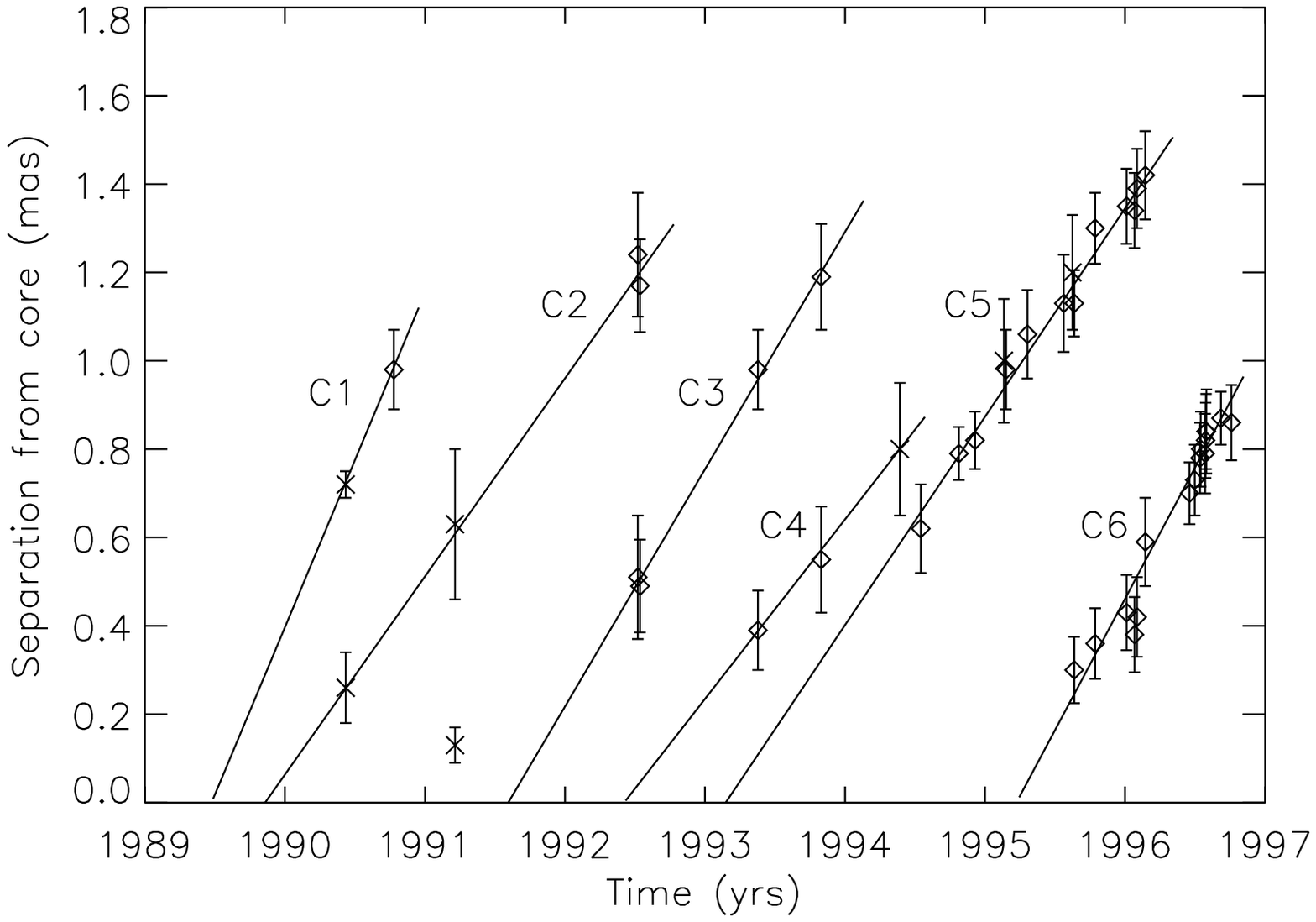}
\caption{Motion of components in OJ 287. The vertical axis shows 
the separation in mas of the center of the component from the core. 
The straight lines provide estimates of the presumed zero separation 
of components C1 to C6 under the assumption  of 
rectilinear motion at constant velocity.
The symbol $\diamond$ corresponds to our data and $\times$ to published data.
At 1990.47 is 8 GHz data from Gabuzda \& Cawthorne (1996), at 1991.27 
43 GHz data from Tateyama et al. (1996), at 1994.52  8 GHz data from 
Fey et al. (1996), and at 1995.28 and 1995.77  8 GHz data from 
Fey \& Charlot (1997).
The error bars on component positions are assumed to be one-eighth
of the beamwidth.}
\end{figure*}

\section{Discussion}

\subsection{Radio Light Curves}

Figure 3 shows the optical and radio light curves of OJ 287.
The beginning of  each radio outburst is indicated by a vertical line.
There is a good agreement between the epochs of the radio outbursts and the 
epochs of
zero separation of
components C1 to C6 which are
1989.5, 1990.0, 1991.7, 1992.5, 1993.3, and 1995.4 respectively. 
Only  C2 would require a higher proper motion for its zero separation epoch to 
match with the beginning
of the radio outburst around 1990.2. However, we may place the 43 GHz  data 
point a little closer
to the core, accounting for a possible frequency-dependent separation between it 
and the 8 GHz data. 
Since the 
8 GHz measurement of Gabuzda \& Cawthorne (1996)
is not clearly resolved from the core, it may also be placed closer to 
the core, resulting in a zero separation epoch for C2 more consistent 
with the beginning of the outburst. 

\begin{figure*}[!t]
\plotone{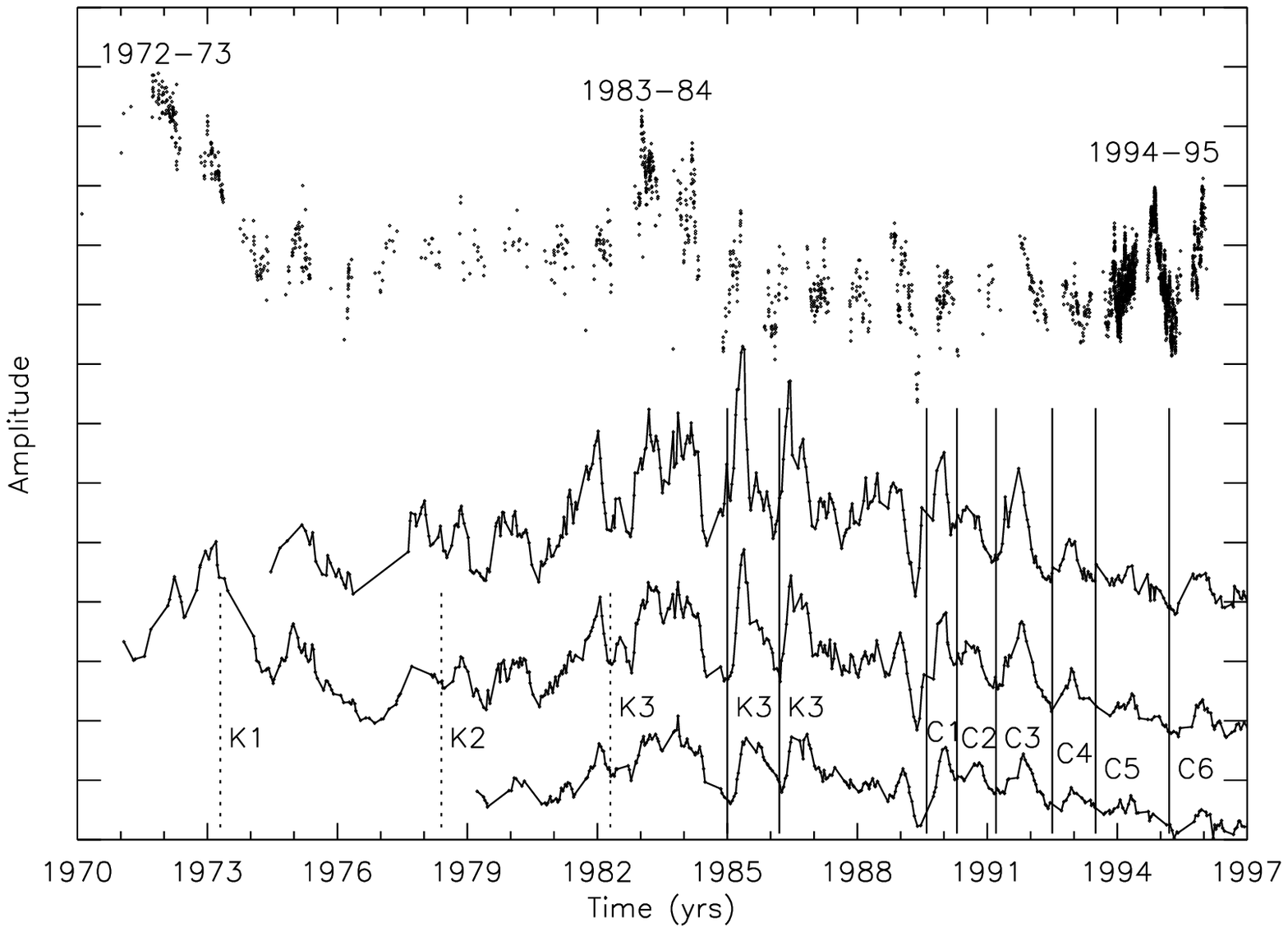}
\caption{ Optical and radio light curves of OJ 287. The optical data
are from Figure 2 of  Sillanp\"a\"a et al. (1996).
The radio data are from the University of
Michigan Radio Astronomy Observatory. 
From  top to bottom the plot shows: V-band optical data, and 15 GHz,  8 GHz,
and 5 GHz radio data. Each tick mark on the ordinate corresponds to a flux 
density
of 2 Jy for the radio data, and an amplitude of  1 magnitude for the optical 
data.
The vertical lines indicate the beginning of the
outburst related to each VLBI component. The dashed lines indicate the epochs of
the birth of components K1, K2, and K3. The multiple lines
shown for component K3 are discussed in the text. 
The epochs of double-peaked optical outbursts are indicated
over each such outburst.}
\end{figure*}

The zero separation epoch of C1 is estimated from only two points. The point
closer to the core was taken from 8 GHz VLBI observations obtained by 
Gabuzda \& Cawthorne (1996). The extrapolation of the fitted motion to zero 
separation agrees well with the start of the radio outburst at 1989.5.
Component C3 would be related to the outburst that started in 1991.5. 
There is also a weak knot observed near the core in 1991.27 at 43 GHz,
which may either be component C3 itself or another component.
The radio light curve shows a substructure
which may be linked with this component, suggesting that it might 
have been an intrinsically weak or short-lived component, and faded or merged 
with C3.
Component C4 has 3 data points;  the farthest point from the core is a 
measurement 
obtained from VLBA   data (Fey,  Clegg,  \& Fomalont 1996). 
This component is associated with a single radio outburst  that started in 
1992.5.
The next component, C5,  has the largest separation from the core, 
and is probably associated with a radio outburst that started in 1993.4.  
Two points from VLBA data at 8 GHz  (Fey \& Charlot 1997) have been included in 
the curve.
The small separation between C4 and C5 raises the question of whether or not 
they
can be regarded as one component.
In the one radio outburst per VLBI component scheme, C4 and C5 are individual 
components which merge later into C5. One weak point to this interpretation is 
the non-detection of  C5 closer to the core in the 1993 December data. 
However, it could be too close to C4 to be discernible as a separate 
component at this time.
If C4 and C5 are separate components,  we would expect a merging of these two 
components
around the end of 1994. In fact,  an increase in the flux density of  component 
C5 in 1994 December 
could be taken as an indication of such an event. 
Finally, component C6 could be related to the outburst that started in 1995.4.

We have also examined a series of geodetic VLBI maps at 8 GHz obtained by 
Vicente et al. (1996).
They have interpreted the structure of OJ 287
in terms of helical motion, with an indication of a small loop at a distance of 
about 0.6 mas
from the core. They  claim a second loop at 2.4 mas is consistent with
VLBI data at 5 GHz (Roberts et al . 1987; Gabuzda et al. 1989). Even so,
they found no evidence of boosting of K3 as would be expected from the 
geometrical
effects of a helically moving feature, and the positions of K2 appeared well
below the calculated helical trajectory.  Even revising the K2 positions upwards
(Vicente et al. 1996), the fit does not improve much.
Now,  if we are allowed to break the component corresponding to epoch 
1986.81 (Figure 3 of Vicente et al. 1996) into 2 subcomponents at 0.4 and 0.8 
mas, a
resolution of 0.5 mas would hardly distinguish between the two
models.  We could then have  rectilinear motion for both
components with a proper motion of about 0.4 mas yr$^{-1}$.
An appealing support for the two rectilinear components is 
the existence of well-defined radio outbursts corresponding to their 
estimated zero-separation epochs, as shown on Figure 3.

We have also looked for VLBI/radio-outburst correlations for the older 
VLBI components K1, K2, and K3 (Roberts et al. 1987).
The derived proper motion for these components (0.20 - 0.28 mas yr$^{-1}$)
is about twice as low as our measured proper motion. This indicates that, in the 
past,
different components may have been regarded as the same component.
The imbalance between the number of knots 
and the number of radio outbursts is also noticeable on Figure 3.  There are at 
least
2 well-defined radio outbursts between the one related
to the birth of K2 in 1978.4 and K3 in 1982.3 (Roberts et al. 1987).  A similar 
occurrence is
also present for component K1, indicating that VLBI components may
have been missed. 

\subsection{Optical Light Curve }

On the top part of Figure 3 is the optical light curve from Sillanp\"a\"a et al. 
(1996).
There is a remarkable similarity  between it and the radio light curves.
Despite the complexity of the structures, most features can be found in both 
emission bands.
The most easily recognizable correlated radio-optical feature
is the double structure that occurred in 1983-84, which had a similar amplitude
in the radio and in the optical. 
The double outbursts of  1971-73 and 1994-95 present a pair of almost identical 
optical flares,
while in the radio these features appear almost as single  outbursts coincident 
with 
the second optical flare. However,
upon more detailed inspection, small protrusions on the radio light curves 
coincident with the 
first flare of the double peak are also present  in 1971 and 1994.
Even weak optical-radio substructures can be pinpointed in the feature beginning 
in  1994.
This can be seen in Figure 4, which shows an expanded section of the light curve 
corresponding
to the period of VLBI observations (1990 to 1997).
It is also clear that radio features are delayed by a few months relative to 
optical features. 
The inclined dashed lines on the figure correspond to a  delay of  0.14 yr.
The smoother profile at lower frequencies and the time delay between optical
and radio bands gives strong support for the synchrotron self-absorption process 
(Melnikov 1998).
We propose that radio-optical structures are self-absorbed synchrotron
sources, and the absence of radio features not correlated with optical features 
may be due to
a more compact synchrotron source.

\begin{figure*}[!t]
\plotone{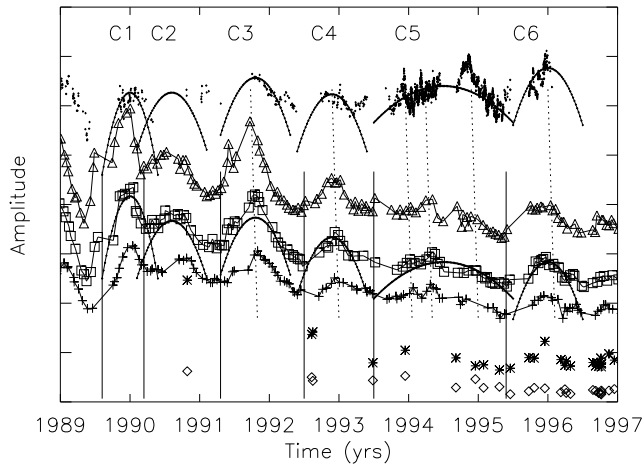}
\caption{ Optical, radio, and VLBI light curves. The ``arcs'' represent 
the size and shape of the outburst associated with each VLBI component. 
They are plotted
on the optical and 8 GHz data. Below the radio light curves the flux of the
core ($\ast$) and the sum of the flux of the jet components ($\diamond$) are 
also shown. 
Except for the optical data, the relative amplitudes are all on the same scale.
Each tick mark on the ordinate corresponds to a flux density
of 2 Jy for the radio data, and an amplitude of  2 magnitudes for the optical 
data.}
\end{figure*}

\subsection{VLBI Light Curves}

Provided that radio and optical emission are synchrotron,  it remains to be seen 
whether this emission
originates in the core, the jet or the accretion disk.
The high density time coverage offered by the geodetic VLBI observations, 
particularly for components C5 and C6, 
enabled us  to study the evolution  of  the flux of the VLBI components 
(core and moving components)  with time as shown on Figure 4 along with the 
radio light curves.
It is clear that the time variation of the flux of the core follows closely the 
shape of the radio light curves,  while the flux of the moving components 
remains nearly constant with time.
With less data, similar results can be seen for  C4 in December 93.
The same behavior can also be seen in geodetic VLBI data studied by Vicente et 
al. (1996), as
shown in Figure 5.
Another aspect of  the flux of the moving components is that
it does not depend on the strength of the radio outbursts.
For instance, a radio outburst  peaking at 7 Jy (K3 of Vicente et al. 1996) 
has an associated  moving component with a flux 
similar to those associated with outbursts  peaking at 2 Jy (e.g., C6 of present 
work).
These  results seem to indicate that the VLBI components  
contribute to the profile of the radio light curve only when they are just 
emerging from the core
at the time of a radio outburst, and are still merged with the core at the 
resolution of these 
VLBI observations.

\begin{figure*}[!t]
\plotone{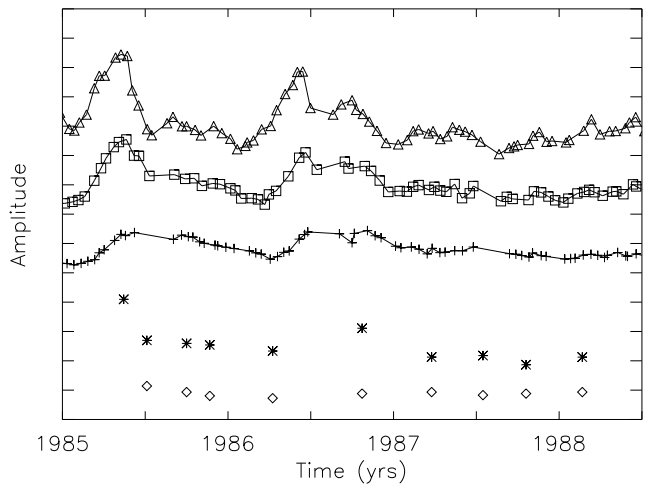}
\caption{ Radio light curves compared to VLBI light curves from Vicente et 
al. (1996).
Below the radio light curves are the flux of the
core ($\ast$) and the sum of  the flux of the jet components ($\diamond$) from 
Vicente et al. (1996).
All amplitudes are on the same scale. Each tick mark on the ordinate corresponds 
to
a flux density of 2 Jy.}
\end{figure*}

\subsection{Binary Systems}

We may speculate that the flares are related to an increase in the accretion 
rate being
induced by the passage of the companion mass  (Sillanp\"a\"a et al. 1988).
We assume that the synchrotron emission is optically thin at optical bands 
and self absorbed at  radio bands.
The absence of  radio outbursts coincident with the first peak of the 
double-peaked optical outbursts
could be an indication of
a more compact synchrotron source. 
We also assume that the major flux variations come from the core.
Keeping in mind a synchrotron emission mechanism for optical and radio 
outbursts;
and knowing  that radio outbursts have irregular periods of about a year 
while major optical flares appear every 12 years, we may conceive a binary model
where the roughly annual outbursts of the core-jet system of the primary are
visited  every 12 years by the secondary, increasing the flux of the outbursts.

A narrow jet placed on the disk axis of either
the coplanar binary model of Sillanp\"a\"a et al. (1988) or the steep inclined 
orbit model
of Lehto \& Valtonen (1996) would provide an appropriate scenario to explain
the observed evolution of the VLBI components as well as the radio
and optical outbursts. Lehto \& Valtonen (1996) have already suggested
a narrow cone on the disk axis to explain a sudden
fading ``eclipse'' caused by the secondary passing over the line of sight.
However,  those binary systems based on the
sweeping beam models of Villata et al. (1996) and Katz (1997)
would be discarded
because they would not account for the observed evolution
of the VLBI components and the number of radio outbursts.
While a binary system through its accretion disk 
increases the accretion rate at evenly spaced intervals, 
the non-periodic nature of nonthermal outbursts would
spread the exact timing of the observable effects of this increased accretion 
rate. 
This is indeed in accordance
with the small variation observed around the period of 12 years.
It would also be possible, depending on the duration of the increased accretion 
rate and the timing 
of the irregular ejection rate, to observe a flare with a 
triple structure as reported by Sundelius et al. (1997).

\section{Conclusions}

We have presented 27 geodetic VLBI maps obtained from the 
Washington VLBI Correlator Facility at the U.S. Naval Observatory. 
These maps showed  a sequence of  6 VLBI components associated with
radio outbursts.
The proper motion of the components was found to be around 0.5 mas yr$^{-1}$,
which is almost twice as high as that seen in previous VLBI observations of this 
source.
Such a higher proper motion, along with the larger number of VLBI
components,  implied a higher component ejection rate
for OJ 287, a result supported by  the close relationship between 
the radio outbursts and the appearance of VLBI components.

We have assumed that 
the radio-optical outbursts are synchrotron emission.
The increase in the accretion rate due to the pericenter passage of the
companion mass would not directly produce 
a VLBI component or radio outburst, but would rather
provide a means to energize the system and 
increase the flux of the synchrotron emission.
The irregular appearance of  radio outbursts/VLBI components 
(about once a year), which is intrinsic to the engine, operates continuously. 
Every 12 years the system is  more apt to produce a higher flux; however,
the ejection rate of VLBI components and
hence the number of radio outbursts is not affected by this process.

\acknowledgments

C.E.T. thanks the grant received  from FAPESP - Funda\c c\~ao de Amparo a
Pesquisa do Estado de S\~ao Paulo (proc. n. 96/6267-1)
to undertake 3 months of work with
geodetic VLBI data at USNO (U.S. Naval Observatory). 
We also thank Aimo Sillanp\"a\"a for supplying the optical data and the referee
for his suggestions and comments which helped with the improvement of the paper.
The University of Michigan Radio Astronomy Observatory is supported by the 
National
Science Foundation and by funds from the University of Michigan.
The Fortaleza VLBI facility was built and is operated with partial support
from U.S. NASA, USNO and NOAA, Brazil ministry of Science and Technology,
MCT-FINEP, Mackenzie, INPE and CRAAE (joint center between Mackenzie,
INPE, USP and UNICAMP, Brazil).

\end{document}